\documentclass[aps,twocolumn,article,nofootinbib, superscriptaddress]{revtex4}
\usepackage{graphicx}
\newcommand{\be}{\begin{eqnarray}}
\newcommand{\ee}{\end{eqnarray}}

\begin{document}

\title{Dominant Folding Pathways of a $\beta$ Hairpin}
\author{Pietro Faccioli\footnote{Corresponding author. Email: faccioli@science.unitn.it}}
\address{Dipartimento di Fisica, Universit\'a degli Studi di Trento, Via Sommarive 14, Povo (Trento) Italy, I-38100.}
\address{I.N.F.N., Gruppo Collegato di Trento, Via Sommarive 14, Povo (Trento) Italy, I-38100.}
\author{Alice Lonardi}
\address{Dipartimento di Fisica, Universit\'a degli Studi di Trento, Via Sommarive 14, Povo (Trento) Italy, I-38100.}
\address{Dipartimento di Scienze e Tecnologie Chimiche, Universit\'a di Roma Tor
Vergata, Via della Ricerca Scientifica I-00133 Rome, Italy}
\author{Henri Orland}
\address{Institut de Physique Th«eorique, Centre dÕEtudes de Saclay, CEA, IPhT, F-91191, Gif-sur-Yvette, France }
\begin{abstract}
We use the Dominant Reaction Pathway (DRP) approach to study the dynamics of the folding of a $\beta$ hairpin, within a  model
which accounts for both native and non-native interactions. We compare the most probable folding pathways calculated with the DRP method with those obtained directly from molecular dynamics (MD) simulations. We find that the two approaches give  completely consistent results. We investigate the effects of the non-native hydrophobic interactions on the folding dynamics found them to be  small. 
\end{abstract}

\maketitle

\section{Introduction}

The theoretical investigation of the microscopic dynamics driving the protein folding reaction remains a very challenging open problem. In general, numerical simulations  based on molecular dynamics (MD) are very inefficient for this purpose. The reason is that most of the 
computational  time is wasted to simulate the thermal oscillations in the (meta)-stable states, while the relevant information about the folding mechanism is encoded in the reactive trajectories which connect denatured and native configurations.

In order to overcome these difficulties, alternative approaches have been developed which yield directly the folding pathways, without investing time in simulating the exploration of metastable states~\cite{TPS1},  \cite{Doniach} \cite{Elber1, Elber2, Elber3} \cite{DRP1,DRP2}. In particular, the DRP method \cite{DRP1, DRP2, DRP3, DRP4} has the advantage to yield directly the \emph{most statistically relevant reaction pathways} in Langevin dynamics, and to sample the reaction using equal configurational displacement steps, rather than equal time steps. This way, it is possible to characterize a folding transition using only a few tens of path discretization steps. 
The DRP approach has been recently extended to predict the dynamics of both electronic  and nuclear degrees of freedom during thermally activated reactions, from quantum mechanical calculations in the Born-Oppenheimer approximation~\cite{QDRP}. 
 
The DRP method is in general very computationally efficient compared to "brute"-force MD simulations. However, there are several potential limitations which have to be taken into account. First,  it cannot predict the structure of the native and denatured configurations, but only determine the dynamics of the conformational transitions  which connects \emph{given} input initial and final configurations. While for many proteins the structure of the native state  has been  accurately  determined from X-ray cristallography or NMR experiments. much less is usually known about the conformational structure of polypeptide chains in the denatured state. 
A commonly adopted strategy to generate model unfolded conformations is to run high temperature  MD unfolding simulations. Such an  approach is certainly appropriate to study the folding of molecules which have been denatured by temperature jumps (see e.g. \cite{eaton} and references therein). 

A second limitation which is common to all approaches which focus on the  reaction pathways~\cite{TPS1}, \cite{Doniach}, \cite{ Elber1,Elber2}, \cite{DRP1,DRP2} is represented by the computational difficulty of performing an exhaustive exploration of the transition path space. Problems arise from the fact that the folding pathways which are sampled
depend on a very large number of degrees of freedom.  In addition, the exploration of the path space is slowed down by the intrinsic ruggedness of the energy landscape. 
In view of such difficulties, the question has been raised whether the folding trajectories obtained in such approaches can be considered realistic representation of the folding reaction.

 \begin{figure}[t!]
	\includegraphics[clip=,width=6 cm]{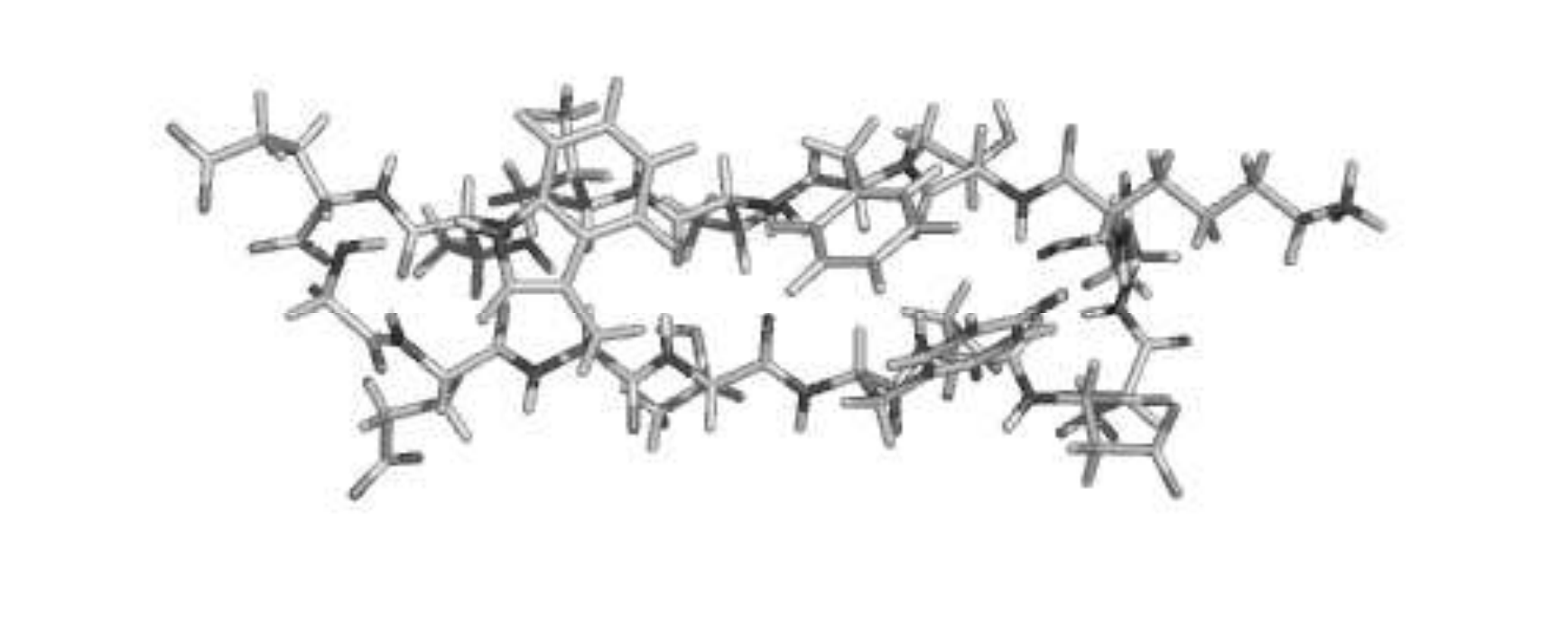}		
	\includegraphics[clip=,width=5 cm]{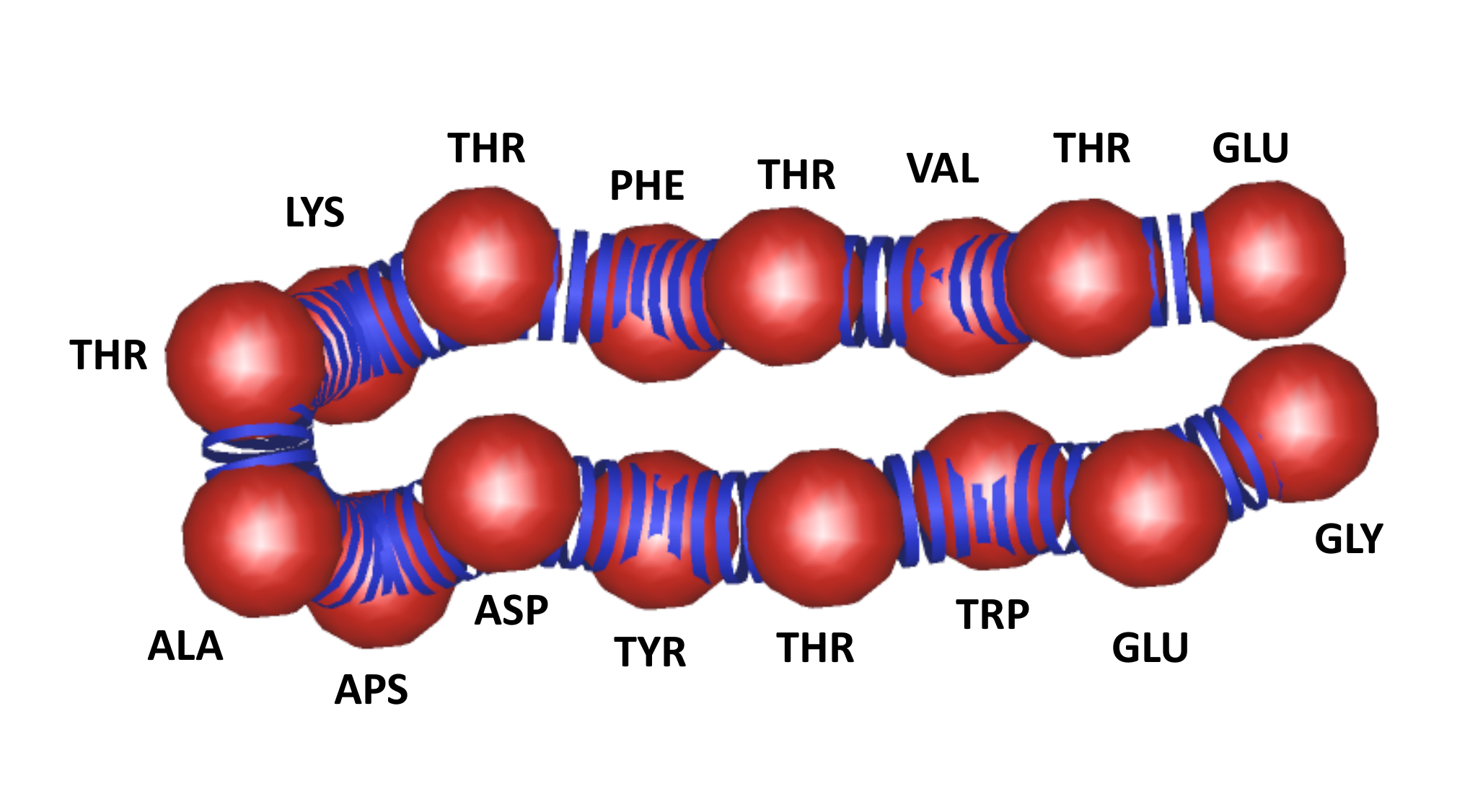}		
	\caption{Upper panel: the 16-residue C-terminus of protein G-B1 (PDB code 2gb1). Lower panel: its coarse grained representation used in the present study.}
	\label{GB1}
\end{figure}

Finally, an important  question which remains to be answered is how many of such reactive trajectories are needed in order to completely characterize the dynamics of a  folding transition. 
Indeed, if the set of different pathways along which a protein can fold is exponentially large, then  any theory which allows to simulate only a relatively small  number  of folding trajectories will be useless. 
In this case, one must necessarily rely on macroscopic descriptions,
based on  the dynamics of suitably chosen reaction coordinates--- see e.g. \cite{wang}---, or on the kinetic transitions between thermodynamic states \cite{markov1, markov2, markov3}. By contrast, if the evolution of arbitrary order parameters during the folding reaction can be accurately inferred  from, say,  a few tens of 
 simulated folding trajectories,  then microscopic approaches based on computing folding trajectories in configuration space become useful. 

In view of  these potential limitations, in \cite{DRP4} a study of the accuracy of the DRP approach in predicting protein folding trajectories was performed using an off-lattice Go-type model of  a  16-residue  polypeptide chain. 
It was shown that, by averaging over a handful of dominant reaction trajectories ---each one corresponding to a different initial  condition in the denatured state---the region of lower free energy connecting the native  and denatured state could be accurately located.

In principle, it is not guaranteed that the DRP approach remains reliable when one considers more sophisticated models, characterized by a higher degree of frustration.
Indeed, more complicated energy functions may significantly increase the space of statistically significant protein folding pathways making it more difficult or even unfeasible to identify the physically important pathways.  
It is therefore important to test the DRP method using models in which the landscape is corrugated. The simplest way to do so is to consider a coarse-grained model in which also non-native interactions are taken into account\footnote{PF thanks V.J. Pande for an important discussion on this point.}.

In this work we implement  a Go-type coarse-grained model for the polypeptide chain considered in \cite{DRP4}, in which non-native interactions are also included,  based on the hydrophobic or polar character of the residues. We compare directly the  folding pathways extracted from long MD simulations,  with the corresponding dominant folding trajectories, obtained in the DRP approach.


 \begin{figure}
	\includegraphics[clip=,width=8 cm]{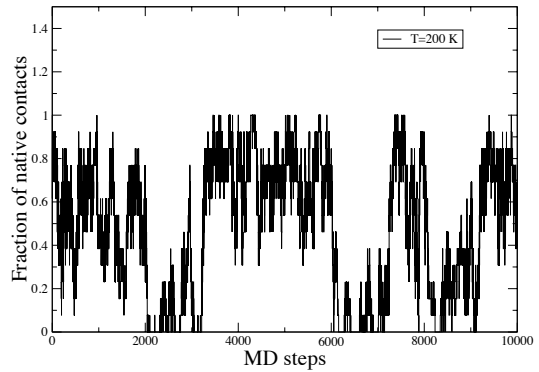}		
	\includegraphics[clip=,width=8 cm]{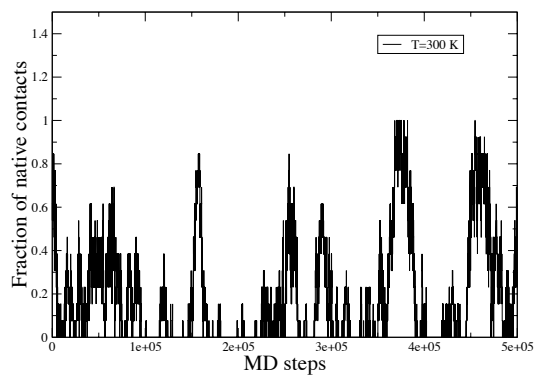}		
	\caption{Time evolution of the fraction of native contacts in a MD simulation based on the coarse-grained model used in this work, at the temperatures $T=200$~K and $T=300$~K.}
	\label{twostate}
\end{figure}

\section{Results and Discussion}
\label{results}

We have used the coarse-grained model defined in section~\ref{model} to  investigate the folding dynamics of the 16-residue 
C-terminus of protein GB1. In the native state of the intact protein, this terminus assumes the structure shown in the upper panel of Fig.~\ref{GB1}. NMR experiments indicate   
that, in aqueous solution and ordinary thermodynamic conditions,  the structure of the hydrophobic cluster is preserved also in the isolated protein terminus\cite{expbeta}. 
Fig. \ref{twostate} shows the time evolution of the fraction of native contacts\footnote{Two residues are considered in contact if their distance is less than 0.6 nm. } obtained from $10^4$ steps of MD simulations (using Brownian dynamics) at the temperatures $T=200$~K and $T=300$~K. These results show that the equilibrium configurations generated by coarse-graiend model form two states:  a native state  (with $x_n\gtrsim0.6$) and a denatured state (with $x_n\lesssim0.3$).  At $T=200$~K ($T=300$~K) the equilibrium population is predominantly native (denatured).
\begin{figure}
 \includegraphics[clip=,width=7cm]{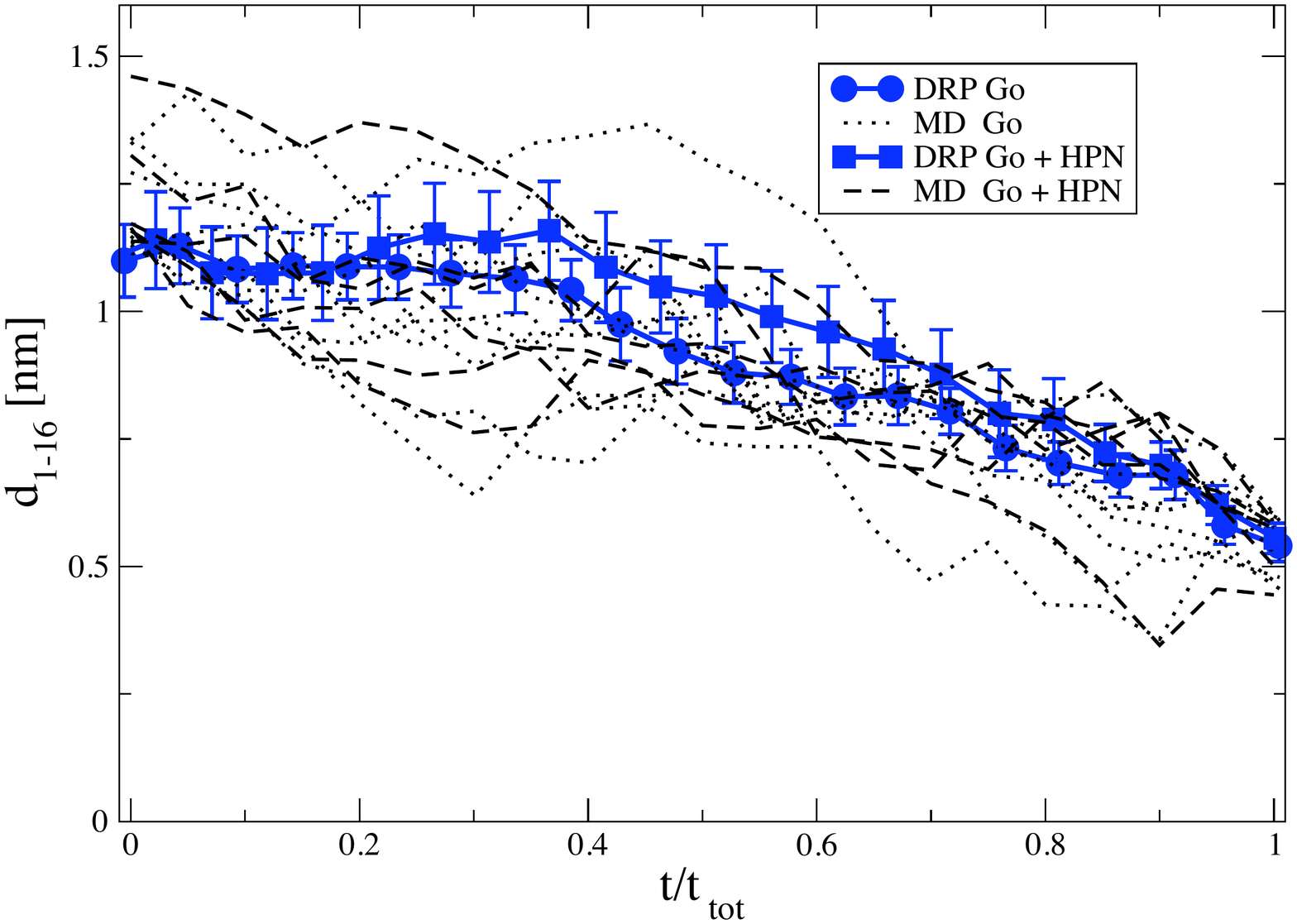}\qquad
  \includegraphics[clip=,width=7 cm]{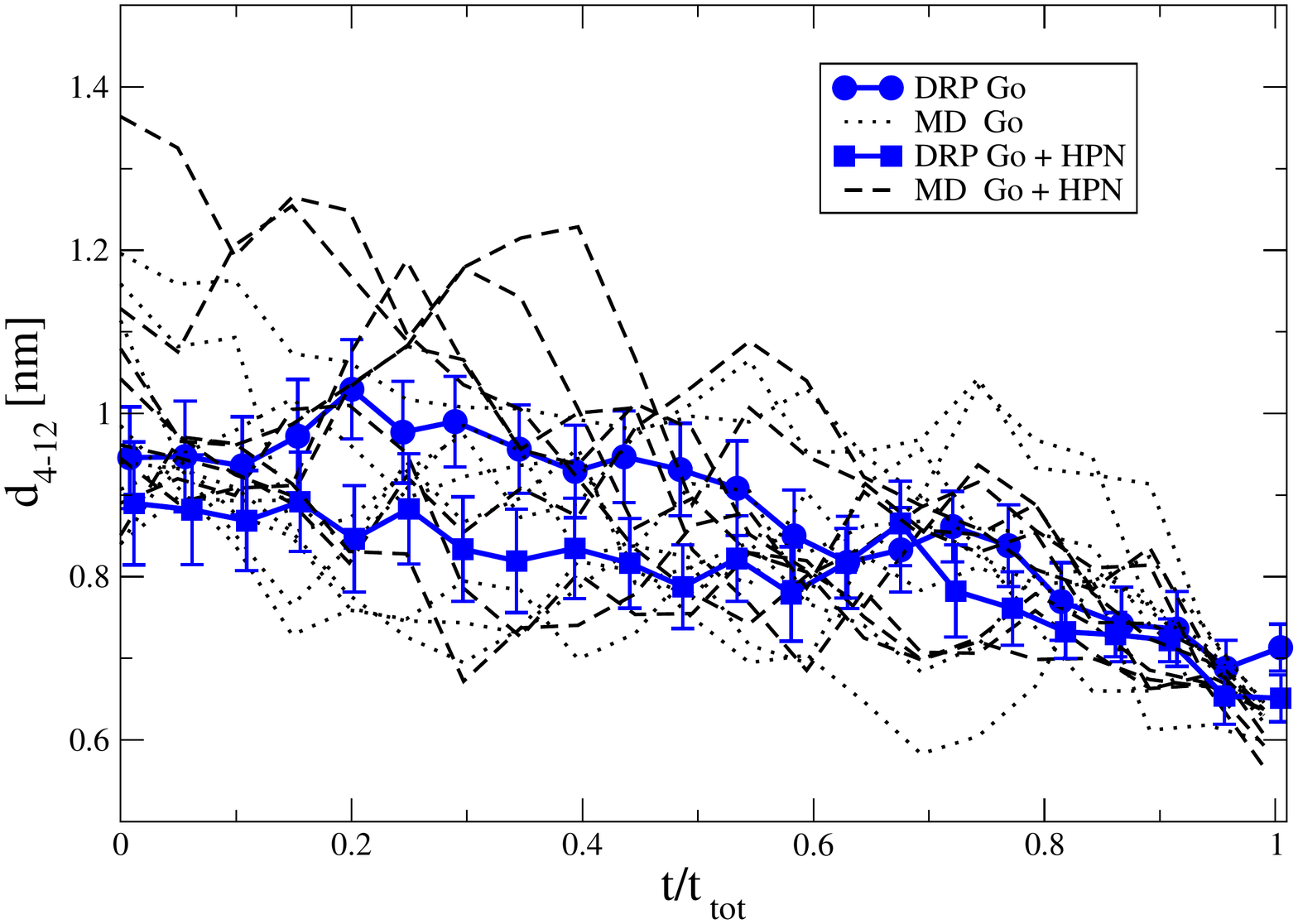}\\
   \includegraphics[clip=,width=7 cm]{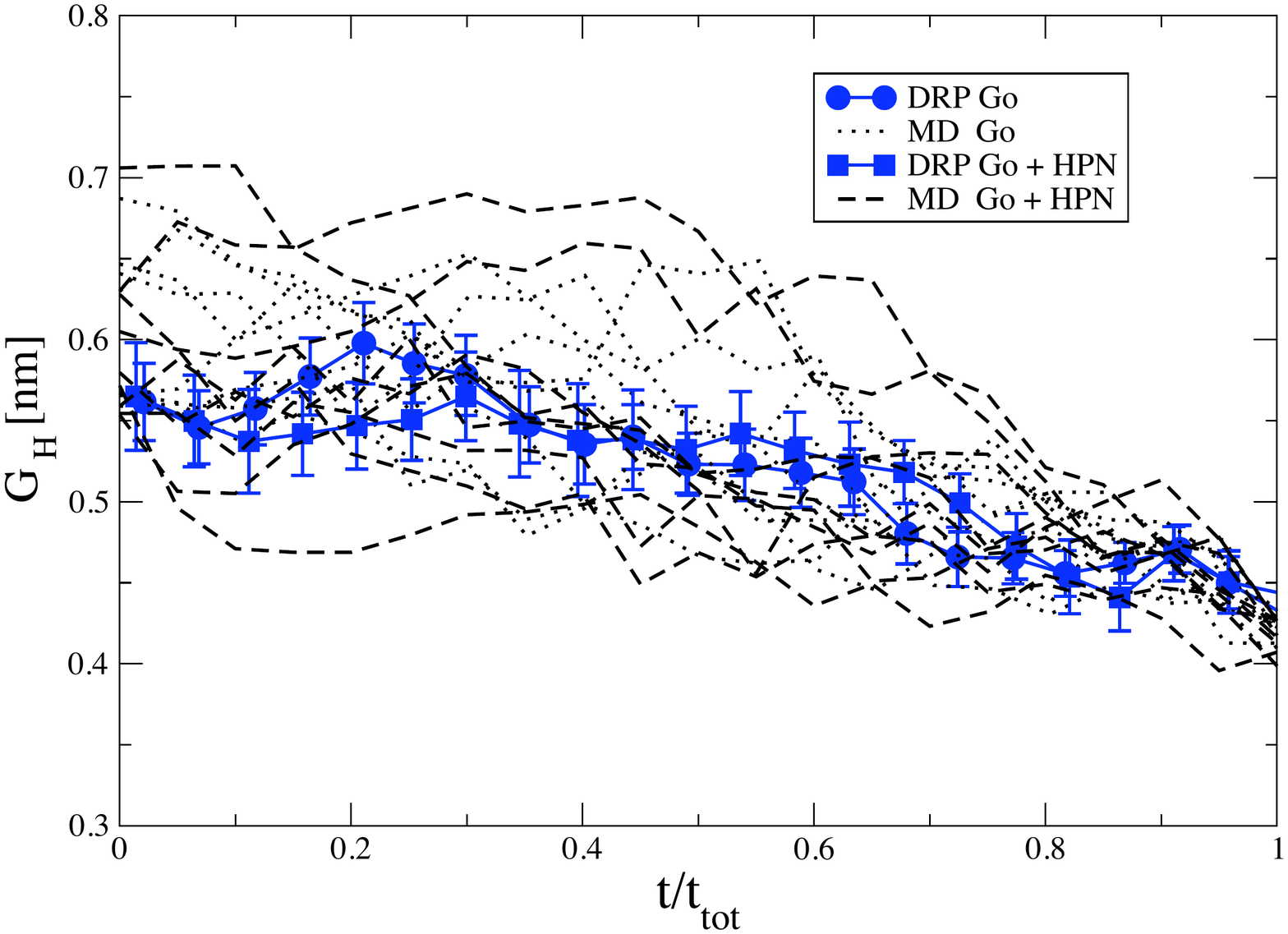}\qquad
   \includegraphics[clip=,width=7 cm]{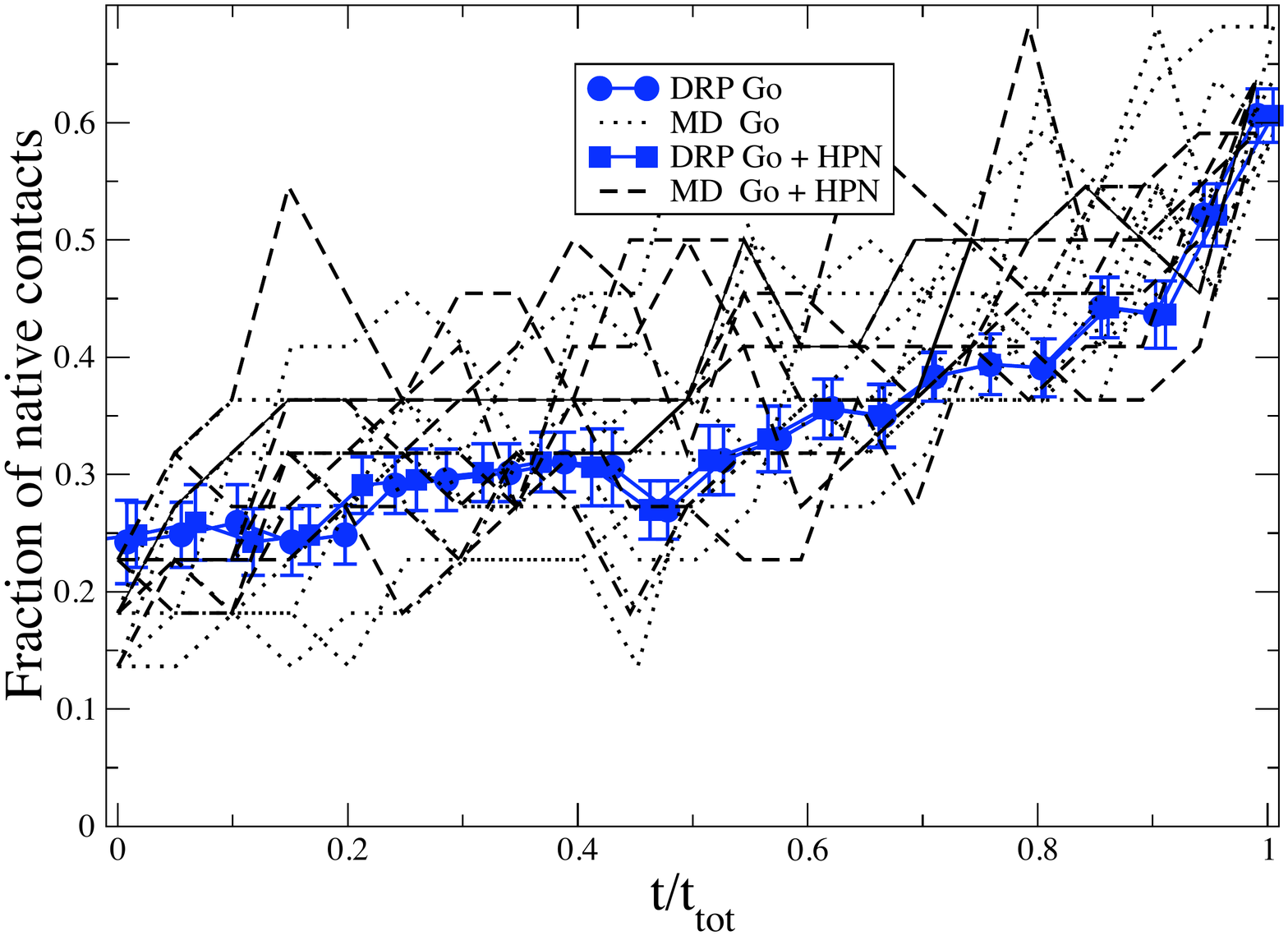}
\caption{Time evolution of the order parameters $d_{1-16}$ (upper panel), $d_{4-12}$ (second  panel from the top) $G_H$ (third panel from the top) and $x_n$ (lowest right) as a function of the fraction of total folding time. Squares and circles denote the results of the DRP with and without non-native contacts, respectively. Dashed and dotted lines represent the results of MD trajectories obtained using a potential energy function with only native contacts and with both native and non-native contacts, respectively.}
\label{res2}
\end{figure}
 
The dominant reaction trajectories calculated with the DRP approach described in section \ref{model} were  used to obtain predictions for the \emph{average} time evolution of observables, using Eq. (\ref{OPI}). In particular, in  this work we considered the dynamics of the following set of order parameters:
\begin{itemize}
\item  The distance between residues 1 and 16, $d_{1-16}(\tau)$: the parameter was defined to be  native for $d_{1-16}< 0.6$nm and denatured  for $d_{1-16}>1.1$nm
\item  The distance between residues 4 and 12, $d_{4-12}(\tau)$: the parameter was defined to be  native for $d_{4-12}< 0.4$nm and denatured  for $d_{4-12}>0.8$nm 
\item The radius of gyration of the hydrophobic cluster formed by the residues  TRP, PHE and TYR, $G_H(\tau)$:  the parameter was defined to be  native for $G_{H}< 0.45$nm and denatured  for $G_{H}>0.55$nm 
\item The fraction of native contacts $x_n(\tau)$: the parameter was defined to be  native for $x_{n}>0.7$ and denatured  for $x_{n}<0.3$.
\end{itemize} 

The results of our DRP calculation at $T=300$~K are presented in Fig.\ref{res2}, where they are compared with the results obtained by evaluating the same order parameters along the folding pathways calculated directly from long MD simulations, in the same model and at the same temperature.  Since the friction coefficient $\gamma$ is a free parameter of the present model, whose only role is to set the total time scale, we plotted the results as a function of  the total transition time $\tau/t$.
The model denoted with "Go" corresponds to one in which only native interactions are included, while "Go-HPN" refers to a model in which non-native hydrophobic and hydrophilic effects are taken into account as well (see section \ref{model} for further details). 

These results show that,  at least for the system under consideration,  the DRP method  quantitatively predicts the average dynamics of all the order parameters considered, during the folding transition. This represents a clear evidence that the method is predictive even if a relatively small number of dominant paths are considered and even if the energy landscape is quite rugged. 

Interestingly, we find that the structure of the folding pathways is not significantly altered, when non-native interactions are removed. A small discrepancy between the two models seems to emerge only for the dynamics of the distance between the two THR residues at position 4 and 12, in the chain (second panel from the top). 
The evolution of such a parameter is the result of the competition between the effective hydrophilic repulsion between these residues, and their tendency to collapse, which is generated by the presence of a hydrophobic cluster near by, formed by the TRP, PHE and TYR residues. In the initial phase of the folding the burying of the hydrophobic cluster overrules the effective hydrophilic repulsion between the THR residues, and the distance $d_{1-12}$ falls shorter in the Go-HPN model than in the simple Go-model. However, in the last stage of the folding the two THR come very close together and the hydrophilic effect tends to keep them more separated than in the Go-model. The crossover between the hydrophobic dominated stage and the hydrophilic dominated stage of the reaction occurs approximatively at two-thirds of the total  folding time. This is an example of the type of dynamical information which is made accessible by the DRP approach. The overall agreement between Go and Go-HPN calculations
represents an evidence in support of the reliability of our analysis of the folding mechanism for this systems, based on Go-type models --- see e.g. \cite{eaton}---.

To summarize, in this work we have tested the accuracy of the DRP method, by comparing directly its predictions for the average time evolution of a set of order parameters during the folding of a $\beta$-hairpin,  with  the folding trajectories obtained directly from molecular dynamics simulations, within in the same model. We found that the two approaches give results which are completely consistent, within the statistical errors. 

We have analyzed the role of non-native interactions in determining the structure of the folding pathways. We  found that in order to accurately predict the time evolution of observables it is sufficient to average over a relatively small set of  dominant paths, each one corresponding to a different initial condition.  Furthermore, we found that the method remains reliable even when the underlying energy landscape generated by the potential energy is quite  corrugated.  

\section{Methods}
\label{model}
\subsection{Definition of the Coarse-Grained Model}

We adopt a coarse-grained representation of such a poly-peptide chain, in which the explicit degrees of freedom are beads which describe the single  amino-acids ---see the lower panel of Fig. \ref{GB1}---. 
The energy function of this model is assumed to be the sum of pair-wise interactions: 
\be
&&U({\bf R}) = \frac{1}{2} \sum_k  k (|{\bf r}_{k+1}-{\bf r}_k|-a)^2 +\nonumber\\
&& \frac 12\sum_{i\ne j} \epsilon \left[A_{ij} \left(\frac{\sigma}{|{\bf r}_{j}-{\bf r}_i|}\right)^{12} - (G_{ij}+B_{ij})~\left(\frac{\sigma}{|{\bf r}_{j}-{\bf r}_i|}\right)^{6}\right]\nonumber\\
\ee
The first term provides chain connectivity, where $a=0.38$~nm represents the average distance between two consecutive $\alpha$-carbons on the chain and $k=3000 ~\textrm{kJ mo}l^{-1} \textrm{nm}^{-2}$ is the elastic constant of the harmonic spring. 
The strength of the Lennard-Jones attraction is set by the parameter $\epsilon= 4 ~\textrm{kJ mol}^{-1}$, while $\sigma=0.3~$nm represents an effective residue size. 
$G_{ij}$ is the matrix of native contacts, i.e. $G_{ij}$  is set to $1$ if the distance between the residues $i$ and $j$ in the native conformation is less than $0.65$~nm, and  $0$ otherwise. The coefficients $A_{ij}$ and $B_{ij}$ introduce residue specificity based on the  hydro-philic and hydro-phobic characters of the individual amino-acids. In analogy with the so-called HP model\cite{HPmodel}, they are defined as follows:
\begin{itemize}
\item  $A_{ij}=1 $ and $B_{ij}=1 $, for pairs in which both amino-acids are  hydrophobic~(H)
\item $A_{ij}=\frac{2}{3}$ and $B_{ij}=-1$,  for  paris in which one of the amino-acids is polar~(P)
\item $A_{ij}=1$ and $B_{ij}=0$ if one of the residues is GLY, which is hydrophobically neutral~(N)
\end{itemize}
 In the following, we shall refer to the model in which the $G_{ij}$,  $A_{ij}$ and $B_{ij}$ coefficients are defined this way as to the Go-HPN model.  Clearly, by setting all $A_{ij}=1$ and $B_{ij}=0$ one recovers the minimally frustrated Go-model used in \cite{DRP4}.

As a concluding remark for this section, we stress that the coupling $B_{ij}$ and $G_{ij}$ do not represent disentangled  physical effects, since the stability of the native structure is  known to be largely influenced by the hydrophobic effect.  Hence, such an effect is encoded implicitly also in the structure of the contact map $G_{ij}$. 
The rationale for adding also the  $B_{ij}$ coefficients to the potential energy is to account for \emph{non-native} hydrophobic interactions and to increase the frustration of the model.
By switching on and off the $B_{ij}$ terms it is possible to increase the ruggedness of the energy landscape, and to study how this affects the structure of the folding pathways.

\subsection{The Dominant Reaction Pathways Approach}
\label{DRPrew}

Let us now briefly review the DRP approach --- for a detailed presentation see e.g. \cite{DRP2}---. Let ${\bf R}$ be a  point in configuration space for the macromolecule under consideration. In this particular case, ${\bf R}$ is defined by the coordinates of all 16 residues ${\bf R} \equiv ({\bf  r}_1, \ldots, {\bf r}_{16})$.
We  assume that the dynamics of the protein in solution  obeys the over-damped Langevin Eq.:
\be
\dot{{\bf R}} &=& -\frac{1}{\gamma} {\bf \nabla} U({\bf R}) + {\bf \eta}(t),
\label{L}
\ee
where $\eta(t)$ are usual Gaussian noise functions, obeying the fluctuation-dissipation relationship,
and $\gamma$ is the viscosity coefficient, which is inversely proportional to the diffusion coefficient $D$,  $\gamma= \frac{1}{\beta D}$ ~(with $\beta=\frac{1}{k_B T}$).
The acceleration term, which appears in the original Langevin Eq., can be shown to be negligible for time scales larger than a fraction of  $ps$.

The time evolution of an arbitrary configuration-dependent observable  $O({\bf R})$   during a folding transition lasting a time $t$  is  given by:
\be
\langle O(\tau) \rangle=&\frac{1}{P(t)}&\int d{\bf R}''\int d{\bf R}'e^{-\frac{\beta\left(U({\bf R}'')-U({\bf R})\right)}{2}} h_D({\bf R}')\nonumber\\
&&\hspace{-2cm} h_N({\bf R}'')\int_{{\bf R}'}^{{\bf R}''}\mathcal{D}{\bf R}~O({\bf R}(\tau)) ~e^{- \beta \int_{0}^{t} d\,\tau'~ \left(\frac{\gamma\,\dot{{\bf R}}^2}{4} + V_{eff}[{\bf R}]\right)}\nonumber\\
\label{OPI}
 \ee
 where $P(t)$ is the (un-normalized) probability to fold in the time interval $t$:
 \be
P(t)&=& \int d{\bf R}''\int d{\bf R}' 
 h_D({\bf R}') h_N({\bf R}'')\,e^{-\frac{\beta\left(U({\bf R}'')-U({\bf R})\right)}{2}}\nonumber\\
 &\times&\int_{{\bf R}'}^{{\bf R}''} 
\mathcal{D}{\bf R}~e^{- \beta \int_{0}^{t} d\,\tau'~ \left(\frac{\gamma\,\dot{{\bf R}}^2}{4} + V_{eff}[{\bf R}]\right)},\nonumber\\
\label{Pt}
\ee
$\tau$ is an intermediate instant during the folding, i.e. $0\le \tau\le t$, and $V_{eff}({\bf R})$ is called the  effective potential, defined as
\be
V_{eff}({\bf R})= \frac{1}{4\gamma } \left((\nabla U({\bf R}))^2 - \frac{2}{\beta} \nabla^2 U({\bf R}) \right).
\label{Veff}
\ee
$h_D({\bf R}')$ and $h_N({\bf R}'")$ in Eq.s (\ref{OPI}) and (\ref{Pt}) are the characteristic functions of the  native and denatured state,
 respectively:  $h_{N(D)}=1$ if ${\bf R}$ is a configuration in the native (denatured) state, and $0$ otherwise.

The DRP approach is based on the saddle-point approximation of the path integrals in (\ref{OPI}). The idea is to consider only the folding trajectories with the largest probability, i.e. those for which the "action" functional
\be
S_{eff} = \int_0^t d\tau' \frac{\gamma}{4} ~\dot{\bf R}^2(\tau')+ V_{eff}[{\bf R}(\tau')]
\ee
 is minimum. 
A dramatic simplification is obtained upon observing that the Eq.s of motion generated by the effective "action" $S_{eff}$ are energy-conserving and time-reversible. 
These properties allows us to switch from the {\it time}-dependent Newtonian description to the {\it energy}-dependent Hamilton-Jacobi (HJ) description. We note that this could not be done at the level of the original Langevin equation.

In the HJ framework, the most probable (or so-called) dominant pathways connecting a \emph{given} denatured configuration ${\bf R}_d$ and a \emph{given} native configuration ${\bf R}_n$ is obtained by minimizing numerically ---e.g. via simulated annealing-- a discretized version of the target function (HJ~functional)
\be
S_{HJ}=\int_{{\bf R}_d}^{{\bf R}_n} dl \sqrt{\left(E_{eff}+V_{eff}[{\bf R}(l)]\right)}, 
\label{SHJ}
\ee 
where $dl$ is an infinitesimal displacement along the path trajectory ---see e.g. \cite{DRP2}  for further details---. 
$E_{eff}$ is a free parameter which determines the total time elapsed during the transition, according to:
\be
\label{time}
t_n-t_d=\int_{{\bf R}_d}^{{\bf R}_n}\,dl \sqrt{\frac{\beta \gamma}{4\left(E_{eff}+V_{eff}[{\bf R}(l)]\right)}}.
\ee 
In \cite{DRP3} it was shown that the longest transition time for a single folding event follows from choosing $E_{eff}=-V_{eff}(x_N)$. However, such a choice generally leads to a very low acceptance rate, when one uses a global minimization algorithm such as simulated annealing to find the dominant paths. The reason is that, for most trial moves, the HJ action becomes complex. To avoid this problem, in this work we consider shorter transition times: we begin the minimization of the HJ action using a very large effective energy, $E_{eff} = V_{eff}({\bf x_N}) + \alpha |V_{eff}[{\bf x_N})|$, with $\alpha = 1000$. The parameter $\alpha$ is then gradually reduced to $\alpha=10$, during the minimization.  

We emphasize that the time interval $t=t_n- t_d$ is not the mean-first passage time from the denatured to the native state. It is the time it takes to fold, once the transition has been initiated. In other words, the DRP formalism is concerned only with the dynamics in the reactive part of the trajectories and not with the dynamics of exploration of the native and denatured states. 

\subsection{Numerical Implementation of the DRP Approach}

\begin{figure}
 \includegraphics[clip=,width=8cm]{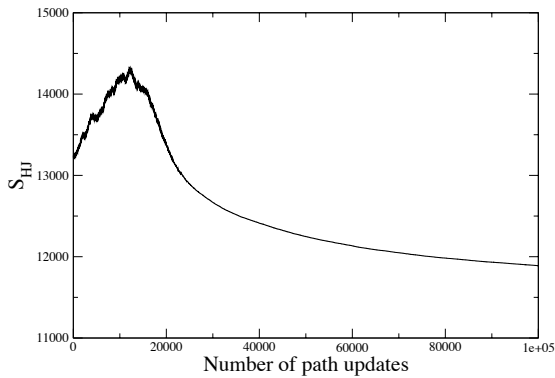}
 \caption{Typical evolution of the HJ action during the numerical minimization based on the simulated annealing algorithm.}
\label{action}
\end{figure}

The numerical minimization of the HJ functional requires the choice of a set of native and denatured configurations.  In addition, one needs to define a corresponding set of trial trajectories connecting the native and denatured configurations, from which the numerical minimization of the HJ functional is initiated. 
A sample of native configurations can be easily obtained  from short MD simulations at room temperature, starting from the  experimentally known native structure. In the present work, we generated the denatured configurations and the corresponding initial trial paths from 30 MD unfolding simulations at $T=2000$~K, starting from the  determined native configurations. The number of frames in such trajectories were decimated
to 23 by averaging the residue coordinates over blocks of consecutive frames, as in \cite{DRP4}.   The search for the dominant reaction pathways starting from
each of such trial trajectories was performed by applying 100,000 steps of simulated annealing.  At each step, the update of the system's configuration
was made according to the following prescription: in each frame of the path, a global trial move was performed using  either global cartesian shifts of all residue coordinates, or by rotating part of the molecule around one of the bonds (i.e. the pivot algorithm~\cite{pivot}). The probability of performing a move based on the pivot algorithm was larger for the first frames --- which correspond to denatured configurations--- while the cartesian moves were statistically favored in the frames corresponding to the latest stage of the folding, where residues are packed and moves based on the pivot algorithm would have a lower acceptance rate. The boldness of the moves was adapted to keep the global acceptance rate around $50\%$.  The behavior of the HJ action during a typical minimization is given in Fig. \ref{action}.

Ideally, the resulting dominant pathways should be independent on the choice of the  initial trial paths. 
In practice, any global minimization algorithm  allows to explore only some functional neighborhood of the initial conditions. Hence, for the DRP to work,  the high-temperature unfolding trajectories  must not be too different from the folding pathways at room temperature. 

\subsection{MD Simulations}

MD simulations were performed by solving the over-damped Langevin Eq. (\ref{L}) in the so-called Ito-Calculus. A sub-part of a long MD trajectory was considered a folding pathway when the value of the order parameters  changed from the denatured to native value in 2000 MD steps or less. This value  corresponds to the typical length of the folding transitions observed in the long MD simulations.


\acknowledgments

We thank M.Sega, S. a Beccara, G. Garberoglio and F.Pederiva for important discussions. This work was motivated by comments made by  A.Szabo.

\end{document}